# Magneto-Transport in MoS$_2$: Phase Coherence, Spin Orbit Scattering and the Hall Factor

Adam T. Neal, Han Liu, Jiangjiang Gu, Peide D. Ye*

School of Electrical and Computer Engineering and Birck Nanotechnology Center, Purdue University, West Lafayette, IN 47907, USA

*Correspondence to: yep@purdue.edu

**Abstract**

We have characterized phase coherence length, spin orbit scattering length, and the Hall factor in n-type MoS$_2$ 2D crystals *via* weak localization measurements and Hall-effect measurements. Weak localization measurements reveal a phase coherence length of ~50 nm at T = 400 mK for a few-layer MoS$_2$ film, decreasing as T$^{-1/2}$ with increased temperatures. Weak localization measurements also allow us, for the first time without optical techniques, to estimate the spin orbit scattering length to be 430 nm, pointing to the potential of MoS$_2$ for spintronics applications. *Via* Hall-effect measurements, we observe a low temperature Hall mobility of 311 cm$^2$/Vs at T = 1 K which decreases as a power law with a characteristic exponent $\gamma$=1.5 from 10 K to 60 K. At room temperature, we observe Hall mobility of 24 cm$^2$/Vs. By determining the Hall factor for MoS$_2$ to be 1.35 at T = 1 K and 2.4 at room temperature, we observe drift mobility of 420 cm$^2$/Vs and 56 cm$^2$/Vs at T = 1 K and room temperature, respectively.

With the recent observation of the indirect to direct bandgap transition in MoS$_2$ [1-2] and the demonstration of MoS$_2$ field effect transistors,[3-5] semiconducting transition metal dichalcogenides (TMDs) have attracted renewed interests in nanotechnology and physics communities. In particular, recent experiments have demonstrated superconductivity[6-7] and valley polarization[8-10] in MoS$_2$, demonstrating the possibility for future spintronics, valleytronics or other novel devices based on MoS$_2$. Much interest is focused on the valence band of monolayer MoS$_2$ due its large spin splitting which results in coupling between the spin and valley;[11-12] however, achieving a p-type contact to MoS$_2$ remains a challenge,[13] limiting strictly electronic MoS$_2$ devices to the conduction band. To further understand n-type MoS$_2$'s potential for novel device applications, we have studied weak localization in few-layer MoS$_2$, allowing, for the first time, an estimate of phase coherence length and spin orbit scattering length in MoS$_2$ which does not require optical techniques. Additionally, we have characterized the mobility and carrier density in multilayer MoS$_2$ *via* the Hall-effect from 1 K up to room temperature. The combination of temperature and gate dependent measurements of the carrier density *via* the Hall-effect has also allowed the Hall factor to be determined in MoS$_2$, for the first time.

**Results and Discussion**

We have characterized weak localization in few-layer MoS$_2$, allowing us to probe the phase coherence, spin coherence, and intervalley scattering in this material. Figure 2 (a) shows the differential magneto-conductance of the sample as a function of magnetic field for different applied back-gate bias at a temperature of 400 mK. The magneto-conductance is calculated from the measured $\rho_{xx}$ and $\rho_{xy}$ by matrix inversion. Over the studied range of gate voltages, we see a positive magneto-conductance at zero magnetic field, characteristic of weak localization. For a lower applied gate bias thus lower electron density and mobility, we see the weak localization dip superimposed on a negative magneto-conductance, which we attribute to the classical $B^2$ dependence of the magneto-resistance with a perpendicular magnetic field. The weak-anti-localization feature is expected to be obtained only at much higher surface electric fields.[7]

To gain insight into the phase coherence length, spin orbit scattering, and intervalley scattering in $MoS_2$, we fit the low-field portion of the magneto-conductance curves using the following equation

$$\Delta\sigma = \sigma(B) - \sigma(B=0) = \alpha \frac{e^2}{4\pi^2\hbar} F\left(\frac{B}{B_\phi}\right) \qquad (1)$$

$$F(z) = \psi\left(\frac{1}{2} + \frac{1}{z}\right) - \ln(z), \qquad B_\phi = \frac{\hbar}{4eL_\phi^2}$$

where $\psi$ is the digamma function, $L_\phi$ the phase coherence length, $e$ the electronic charge, $\hbar$ the reduced Plank's constant, $B$ the magnetic field, and $\alpha$ an empirical fitting parameter.[14-17] Other than fundamental constants, $\alpha$ and $L_\phi$ completely determine the shape of the weak localization peak. Note that our definition of $\alpha$ differs from the references[15-17] by a factor of $n_s \times n_v$ because we have removed implied spin and valley degeneracy from Equation 1. We have limited our fitting to magnetic fields less than 600 mT to avoid effects of the classical magneto-resistance at low carrier densities. Examples of curves fitted using this equation are pictured in Figure 2 (a) for $V_{bg} = 100$ V and $V_{bg} = 70$ V at T = 400 mK with solid orange curves. Figure 2 (b) shows the estimated phase coherence length as a function of carrier density, which we find to be about 50 nm over the range studied. A moving average over carrier density was applied to the phase coherence lengths, with the standard deviation indicated by the error bar.

To qualitatively understand the effects of spin orbit scattering and intervalley scattering in MoS$_2$ from the measurements, we consider the empirical fitting parameter $\alpha$ with the results from the fitting shown in Figure 3 (a). We observe that $\alpha$ is positive, but less than two, which we attribute to a combination of strong intervalley scattering and weak spin scattering in the system, relative to the dephasing of the electrons. The value that $\alpha$ takes is determined by the number of independent conducting channels in the system.[16,17] For few-layer MoS$_2$, in the absence of intervalley scattering or spin orbit scattering, $\alpha = 12$, the number of spins (two) times the number of valleys (six) in the system. Once there is intervalley scattering or spin orbit scattering, these conducting channels become coupled and no longer contribute to the weak localization correction independently, leading to a reduction in $\alpha$. In the case of spin orbit scattering, $\alpha$ also undergoes a sign change due to the $\pi$ phase difference between wave functions of opposite spin.[18] Therefore, if there is no spin scattering in the system, $\alpha$ should decrease from 12 to two as intervalley scattering increases. This situation is pictured in Figure 3 (a) as the white to light blue color gradient. As shown in Figure 3 (a), the value of $\alpha$ determined from fitting the experimental data is less than two in general, so the observed shape of the weak localization cannot be explained by intervalley scattering alone. Therefore, we attribute the further reduction of $\alpha$ below two to spin orbit scattering which is strong enough to reduce $\alpha$ but weak enough such that weak anti-localization is not observed.

Because intervalley scattering and spin orbit scattering are simultaneously present in the n-type MoS$_2$ system, separating their contributions to $\alpha$ is not straightforward. This situation is in great contrast to the silicon inversion layer, in which only intervalley scattering is important with negligible spin orbit scattering, or narrow bandgap III-V quantum wells, where only spin orbit scattering is important with a single valley. MoS$_2$ and other TMDs provide a new type of unique material system in terms of studying the interaction of these scattering mechanisms. Although there has been a theoretical treatment of weak localization in the spin-valley coupled MoS$_2$ valence band,[19] the situation in the conduction band, where the spin and valley are not coupled, has not been quantitatively addressed. However, if we assume that the intervalley scattering length is much shorter than the phase coherence length, then we can use established theory to estimate an upper limit on the spin orbit scattering length in this system. We use the following equation[18] to estimate the upper limit of the spin orbit scattering length

$$\Delta \sigma = n_s \frac{e^2}{4\pi^2 \hbar} \left( F\left(\frac{B}{B_\phi + B_{so}}\right) + \frac{-1}{n_s}\left(F\left(\frac{B}{B_\phi}\right) - F\left(\frac{B}{B_\phi + 2B_{so}}\right)\right)\right) \quad (2)$$

$$F(z) = \psi\left(\frac{1}{2} + \frac{1}{z}\right) - \ln(z), \quad B_* = \frac{\hbar}{4eL_*^2}, \quad * = \phi, so$$

where $n_s = 2$ for the spin degeneracy, $L_{so}$ is the spin orbit scattering length, and the other quantities are defined as in Equation 1. Both the phase coherence length and spin orbit scattering length were allowed to vary in this fitting. To avoid effects of the classical magneto-resistance, fitting was limited to magnetic fields less than 600 mT for carrier densities less than $8.6 \times 10^{12}$ cm$^{-2}$. For carrier densities from $8.6 \times 10^{12}$ cm$^{-2}$ to $1.2 \times 10^{13}$ cm$^{-2}$, this upper limit on magnetic field was increased linearly from 600 mT to 1 T.

The estimated spin orbit scattering lengths resulting from this fitting are shown in Figure 3 (b). Again, a moving average over carrier density was applied to the extracted spin orbit scattering lengths. We see that the upper estimate of the spin orbit scattering length increases with increasing carrier density, reaching as high as 430 nm at a carrier density of $1.1 \times 10^{13}$ cm$^{-2}$. We emphasize that, strictly speaking, we can only consider this value as an upper limit on the spin orbit scattering length. If intervalley scattering is weaker than we assume, then our assumption of strong intervalley scattering attributes too much of the reduction of $\alpha$ to intervalley scattering, leading to an overestimated spin orbit scattering length. However, if we assume the opposite extreme, that there is no intervalley scattering in the system, we cannot fit the data by an equation analogous to Equation 2. (Equation 2 would be multiplied by six, the valley degeneracy, in the case of no intervalley scattering.) Therefore, we argue that the spin orbit scatting length determined by fitting the data with Equation 2 is still a useful estimate because there must be intervalley scattering at least comparable to the dephasing present in the system.

The positive α in Fig. 3(a) indicates the weak spin-orbit interaction in the $MoS_2$ films at the measured electron densities. To realize strong spin-orbit interaction in $MoS_2$ films by strong external electric fields and the Rashba effect, it requires gate induced carrier densities an order of magnitude higher than those we achieved here, enabled by ionic liquid gate.[7] In contrast to our experiment as shown in ref.7, the negative α was observed with strong spin-orbit interaction. In this experiment, the increase in spin orbit scattering length is attributed mainly to the increase of the diffusion coefficient with increasing carrier density, because spin-orbit interaction does not dominate. As shown in Supporting Information Figure S2, the mobility, hence the diffusion coefficient and spin-orbit scattering length, increases with increasing back-gate bias and electron density. In general, it is not surprised to observe a relatively long spin orbit scattering length with a weak spin-orbit interaction.

Figure 4 (a) shows weak localization characterized as a function of temperature at $V_{bg}$=100 V. Data for $V_{bg}$=70 V can be found in Supporting Information Figure S1. Fitting again with Equation 1, this allows us to determine the phase coherence length as a function of temperature, shown in Figure 4 (b). The phase coherence length decays as $T^{-1/2}$, which suggests that electron-electron scattering is responsible for the dephasing of the electron wave function.

In addition to characterizing spin coherence, intervalley scattering, and phase coherence in $MoS_2$, we have studied the mobility in $MoS_2$ *via* the Hall-effect as a function of temperature and applied gate bias. The Hall mobility of few-layer $MoS_2$ as a function of back-gate voltage for temperatures ranging from 1 K to 61 K is systematically measured (see Supporting Information Figure S2). Hall mobility increases with increasing electron concentration, reaching a value of 311 cm$^2$/Vs at T = 1 K and an electron concentration of $1.2\times10^{13}$ cm$^{-2}$, the highest mobility measured for these devices over the studied range of experimental parameters. Figure 5 shows a summary of Hall mobility *versus* temperature for different gate bias, thus carrier density. For lower applied gate voltage, mobility increases with temperature from 1 K to 15 K while it decreases as a power law from 15 K to 60 K. As gate bias is increased, mobility decreases as a power law with temperature over the entire temperature range.

Figure 5 (b) shows the characteristic exponent γ describing the power-law decrease of mobility with temperature for the data in Figure 5 (a). Example fits to the reciprocal of the mobility used to extract γ are presented in Supporting Information Figure S3. The scattering mechanism responsible for the observed γ is not completely clear. Theoretical calculations show that intrinsic phonon mobilities for $MoS_2$ are expected to be much higher[20-22] than those experimentally measured, so it seems unlikely that phonons intrinsic to the $MoS_2$ are responsible for this temperature dependence. This fact leads one to consider if phonons from the $SiO_2$ substrate are responsible for the observed temperature dependence, but this also seems unlikely. The lowest surface polar optical phonon energy for $SiO_2$, about 60 meV,[23] is at least an order of magnitude larger than the thermal energy, *kT*, in this temperature range, leading to negligible optical phonon population. Additionally, at our highest doping of $1.2\times10^{13}$ cm$^{-2}$, noting that there are six conduction band valleys in few-layer MoS$_2$, we estimate that the Fermi level, relative to the bottom of the conduct band, is about 10 meV, insufficient for spontaneous optical phonon emission. Therefore, the experimental data suggest that film bulk defects, domain boundaries and other scattering mechanisms should be considered for the MoS$_2$ system in this low temperature regime.[21]

Keeping the potential device applications in mind for MoS$_2$, accurate determination of transport mobility of MoS$_2$ at room temperature is important.[24] Most work is focused on field-effect mobility with two-terminal source/drain configuration and significant contact resistance.[4,5,13,25,26] To more accurately characterize mobility in MoS$_2$, the Hall mobility as a function of gate voltage and temperatures from 110 K to room temperature for another two multilayer MoS$_2$ devices (Device 2 and Device 3), different from the previous one presented in Figures 2-5, are systematically studied. As shown in Supporting Information Figure S4, we see again in this Device 2 that mobility increases with increasing gate bias at 110K, which we associate with decreased Coulomb scattering at higher carrier densities as in Supporting Information Figure S2. At 290 K, however, the mobility quickly saturates with increasing gate voltage and carrier density. Figure 6 (a) shows the temperature dependence of the Hall mobility and carrier density for this device, and at room temperature we observe a Hall mobility of 24 cm$^2$/Vs at a back-gate bias of 30 V and a carrier density of $7.7\times10^{12}$ cm$^{-2}$.

Although the Hall-effect provides a measurement of the carrier density in the system independent of assumptions about the gate capacitance, care must still be taken when comparing Hall mobility to field-effect mobility due to the Hall factor. This Hall factor, if it differs from one, causes a person to overestimate the true carrier density and underestimate the true drift mobility by this factor. The combination of temperature dependent and gate dependent measurements of the carrier density, shown in Figure 6 (b), reveal this effect in $MoS_2$. Figure 6 (b) shows the carrier density determined by the Hall-effect as a function of applied gate voltage for two temperatures and two devices (Device 2 and Device 3). We observe a change in the slope of the carrier density *versus* gate voltage curve when temperature is changed from 110 K to 290 K. This slope is expected to be $r_h C_{gate}/q$, where $r_h$ is the Hall factor and q is charge of the electron. Since the gate capacitance is expected to be constant with temperature from 110 K to 290 K, we can attribute the change in the slope to the temperature dependence of $r_h$.

Given the fact that the $MoS_2$ flake operates in accumulation, the gate oxide thickness is an order of magnitude larger than the channel thickness, and there is no top gate on these devices, an estimate of $C_{gate}=C_{ox}$ from the oxide thickness can allow us to make a quantitative estimation of the Hall factor for $MoS_2$. With this assumption, the slope of the curves in Figure 6 (b) should be equal to $r_h C_{ox}/q$. Using this result, we determine $r_h$ to be 1.43 at 110 K, increasing to 2.32 at 290 K for Device 2 and 1.45 at 110 K, increasing to 2.48 at 290 K for Device 3. Using this Hall factor, we can determine the true drift mobility at room temperature as $r_h \mu_h$, which is 56 $cm^2/Vs$, as opposed to the previously mentioned Hall mobility of 24 $cm^2/Vs$, for Device 2. Additionally, using the same methods just described, we determine that the Hall factor for Device 1 at T = 1 K is about 1.35, which yields a drift mobility of 420 $cm^2/Vs$ from the Hall mobility of 311 $cm^2/Vs$. The drift mobility determined here establish the upper limits of the two-terminal field-effect mobility of these flakes if perfect, zero resistance contacts could be achieved. The accurately determined drift mobility of 56 $cm^2/Vs$ at room temperature for $MoS_2$ crystals without surface passivation provides the baseline to benchmark the processing techniques of chemical vapor deposited $MoS_2$ films and high-k dielectric passivation.

**Conclusions**

In conclusion, we have characterized the Hall factor, Hall mobility, phase coherence length, and spin orbit scattering length in n-type MoS$_2$ via Hall-effect and weak localization measurements. Further analysis of the weak localization shows that an estimate of an upper limit for the spin orbit scattering length is as high as 430 nm in MoS$_2$. We observe a Hall mobility in MoS$_2$ crystals of 311 cm$^2$/Vs at 1 K and 24 cm$^2$/Vs at 290 K. For the first time, we determine the Hall factor for MoS$_2$ to be 1.35 at T = 1 K and 2.4 at T = 290 K, allowing us to obtain the true drift mobility of 420 cm$^2$/Vs and 56 cm$^2$/Vs at T = 1 K and T = 290 K, respectively. The accurately determined room temperature drift mobility serves as a key indicator of MoS$_2$ performance in device applications.

**Methods**

MoS$_2$ flakes were exfoliated onto 300 nm thermally grown SiO$_2$ on heavily doped Si substrate, which was used as a back-gate. An electron beam lithography and lift-off process was used to define Ni/Au 30 nm/50 nm electrical contacts to the MoS$_2$ flakes. Figure 1 shows the schematic picture of the fabricated MoS$_2$ Hall devices. Device 1 has 4-layer MoS$_2$ determined by the optical contrast (see Supporting Information Figure S5) and an atomic force microscopic measurement (see Supporting Information Figure S6). Device 2 and Device 3 are multi-layer MoS$_2$ films with a thickness of 5-6 nm (see Supporting Information Figure S6). Measurements less than T = 70 K were carried out with a superconducting magnet with He$^3$ insert while measurements at higher temperatures were carried out in vacuum using an Abbess Hall-effect measurement system with a resistive magnet. Standard low-frequency lock-in techniques were used for electrical characterization.

*Acknowledgements:* This material is based upon work partly supported by NSF under Grant CMMI-1120577 and SRC under Tasks 2362 and 2396. A portion of low temperature measurements was performed at the National High Magnetic Field Laboratory, which is supported by National Science Foundation Cooperative Agreement No. DMR-1157490, the State of Florida, and the U.S. Department of Energy. The authors thank E. Palm, T. Murphy, J.-H. Park, and G. Jones for experimental assistance.

*Supporting Information Available:* Supporting figures and additional comments. This material is available free of charge *via* the Internet at http://pubs.acs.org

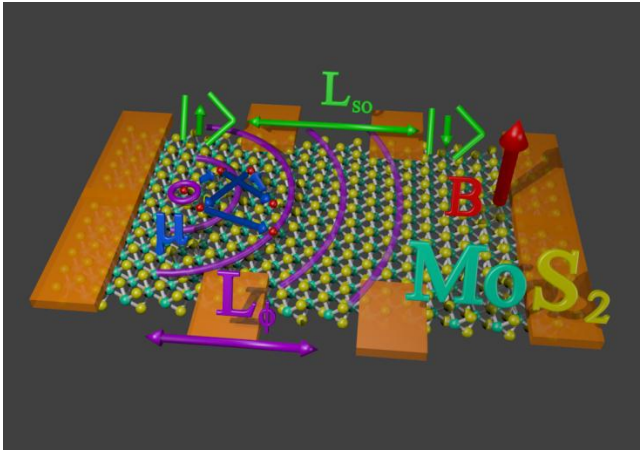

Figure 1: Illustration of the Hall-device structure

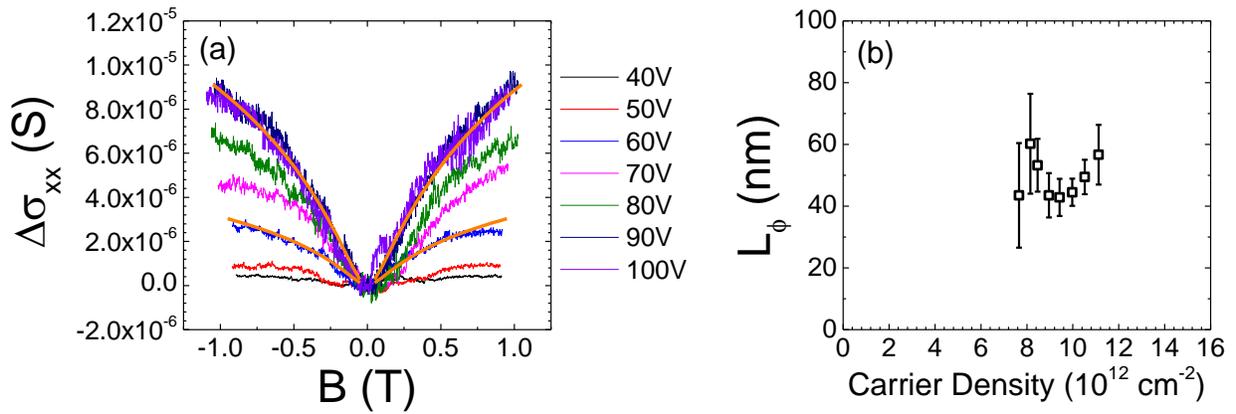

Figure 2: (a) Differential magneto-conductance *versus* magnetic field of electron doped few-layer MoS$_2$ for different applied back-gate voltage. Solid orange lies are example fits using Equation 1. (b) Phase coherence length as a function of carrier density determined by fitting the data in (a) and others not shown using Equation 1. A moving average over carrier density was applied to the data, with the standard deviation indicated by the error bars.

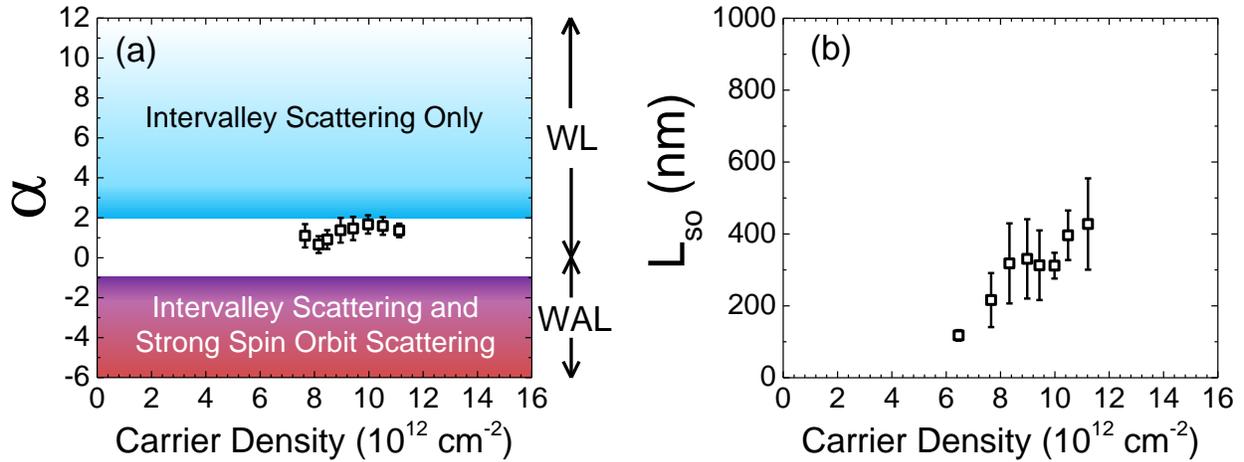

Figure 3: (a) Fitting parameter α as a function of carrier density determined by fitting the data in Figure 2 (a) and others not shown using Equation 1. (b) Upper estimate of the spin orbit scattering length as a function of carrier density from fits using Equation 2. A moving average over carrier density was applied to the data, with the standard deviation indicated by the error bars for both (a) and (b).

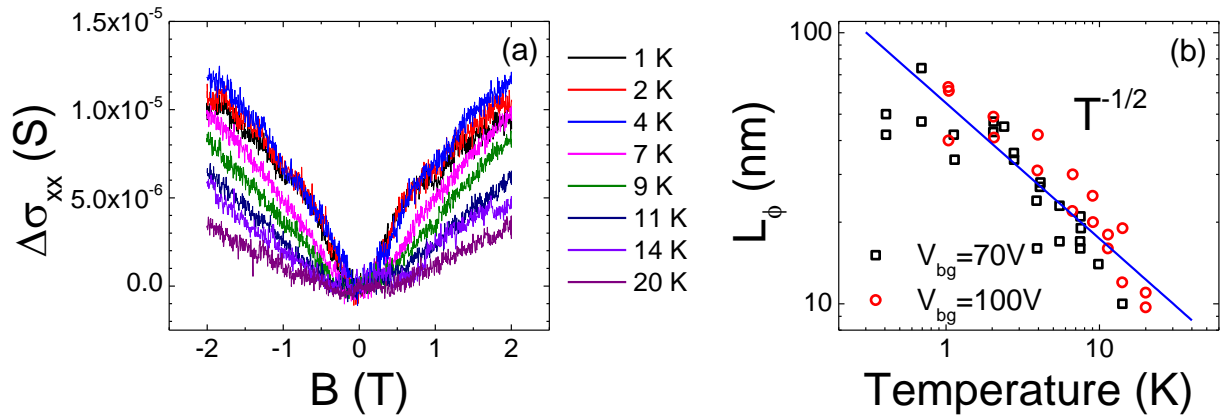

Figure 4: (a) Differential magneto-conductance as a function of magnetic field for electron doped few-layer $MoS_2$ for different temperatures at $V_{bg} = 100$ V (b) Coherence length as a function of temperature from fits using Equation 1.

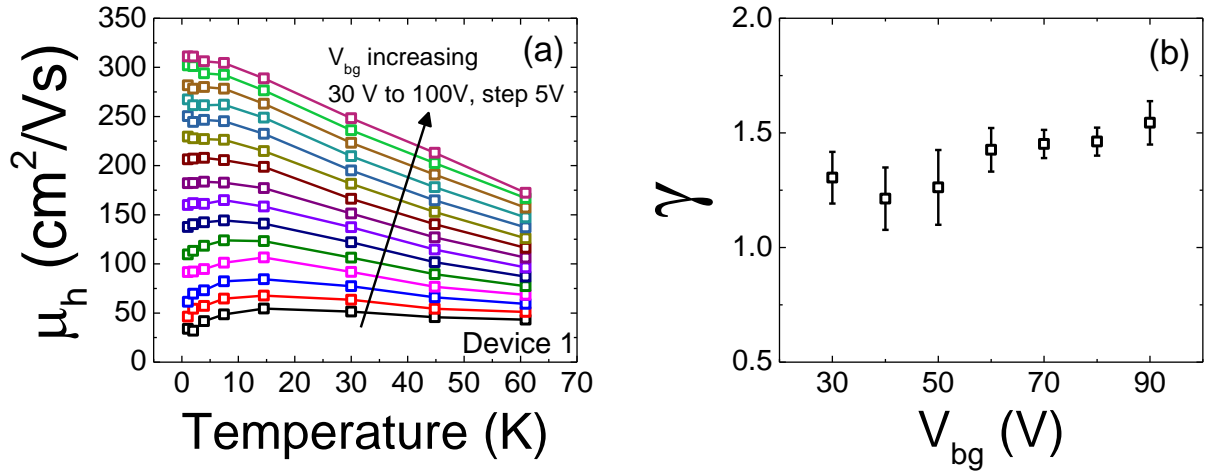

Figure 5: (a) Hall mobility of multilayer MoS$_2$ as a function of temperature. (b) Characteristic exponent of the power law dependence of mobility with temperature extracted from the example fits shown in Supporting Information Figure S3 along with others not shown. A moving average over gate voltage was applied to the data, with the standard deviation indicated by the error

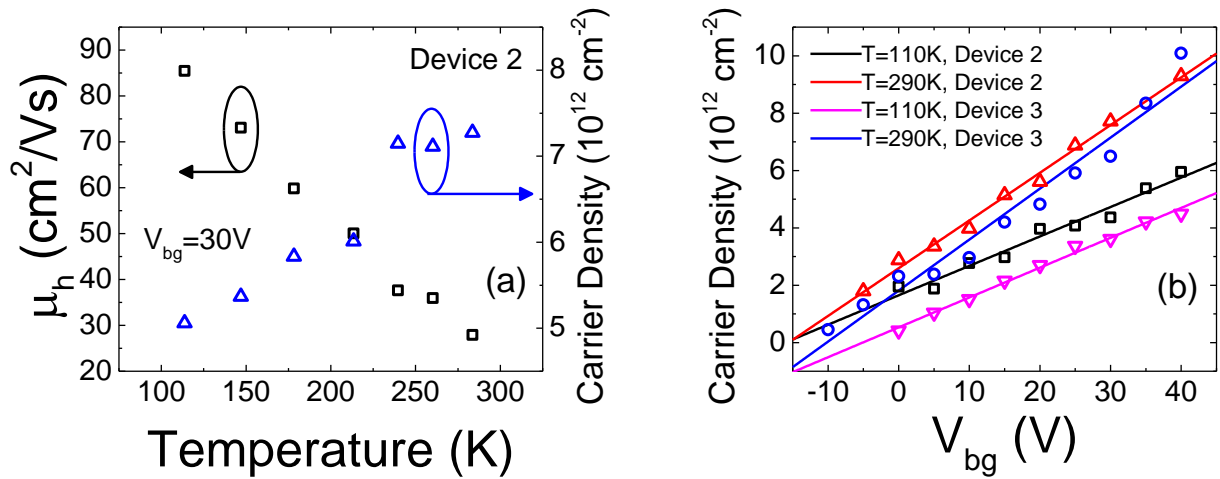

Figure 6: (a) Hall mobility of multilayer MoS$_2$ as a function of temperature. Device is different than that considered in Figures 2-5. (b) Carrier density as a function of gate voltage determined by the Hall-effect for two flakes at T = 110 K and T = 290 K. Symbols are the measured data while solid lines are linear fits. The change in the slope with temperature is due to the change in the Hall factor with temperature.

Table of Contents Figure

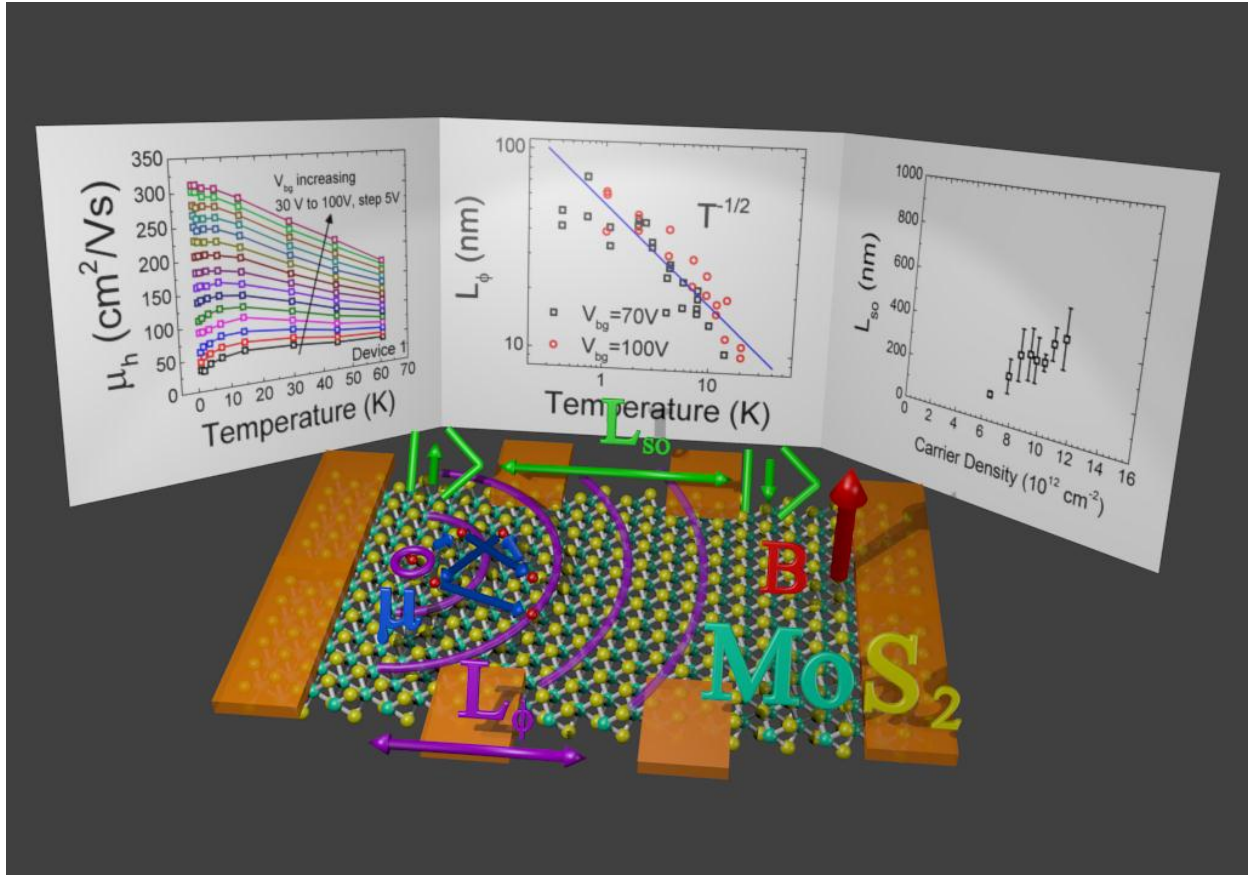

# Supporting Information

# Magneto-Transport in MoS$_2$: Phase Coherence, Spin Orbit Scattering and the Hall Factor

Adam T. Neal, Han Liu, Jiangjiang Gu, Peide D. Ye*

School of Electrical and Computer Engineering and Birck Nanotechnology Center, Purdue University, West Lafayette, IN 47907, USA

*Correspondence to: yep@purdue.edu

**Weak Localization at different temperatures for $V_{bg}$ = 70 V**

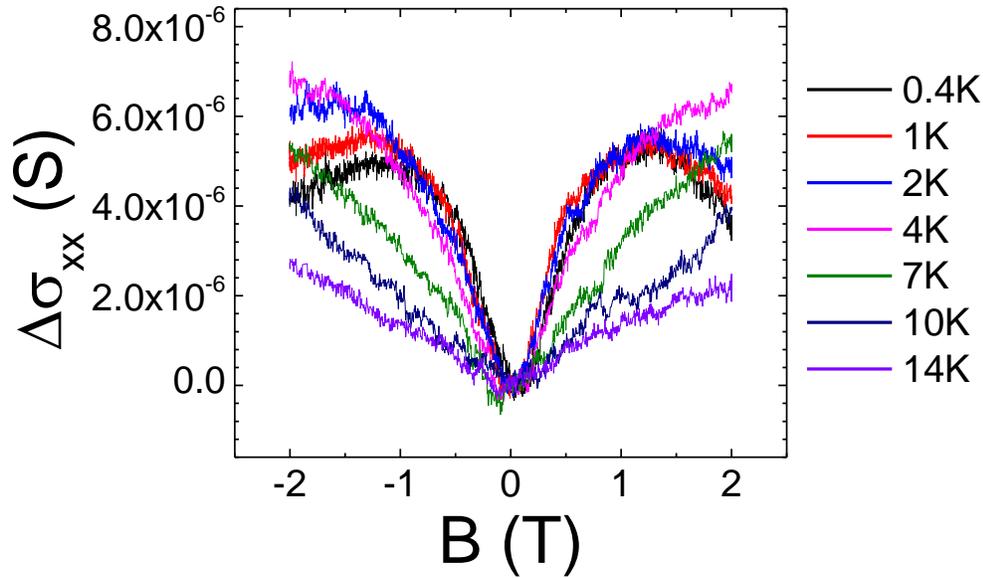

Figure S1: Differential magneto-conductance as a function of magnetic field for electron doped few layer $MoS_2$ for different temperature at $V_{bg}$ = 70 V. This data and others not shown are fit using Equation 1 of the main paper to determine the phase coherence length as a function of temperature at $V_{bg}$ = 70 V. Phase coherence length verses temperature determined from described fitting are plotted in Figure 3 (b) of the main paper along with corresponding phase coherence length data for $V_{bg}$ = 100V. Differential magneto-conductance verses magnetic field data for $V_{bg}$ = 100V is presented in Figure 3 (a) of the main paper.

**Hall Mobility at Low Temperatures in MoS$_2$, T = 1 K to 61 K**

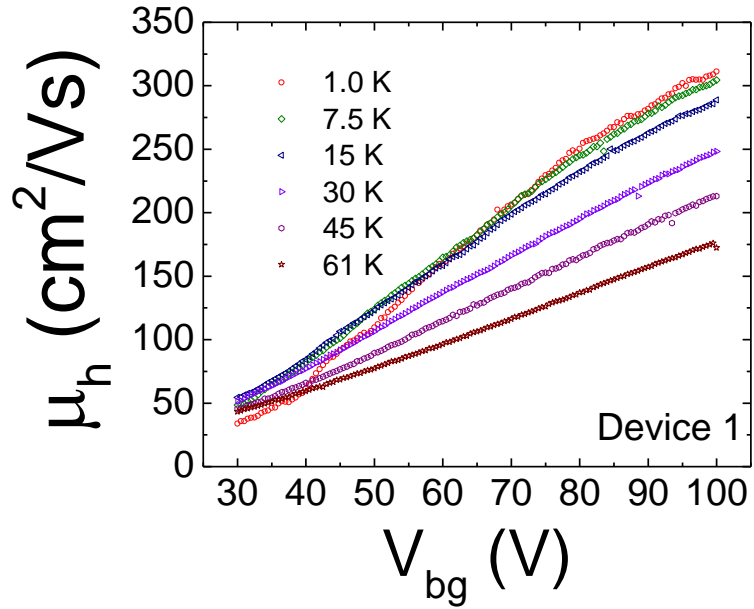

Figure S2: Hall mobility of few-layer MoS$_2$ as a function of back gate voltage for Device 1. The increase in mobility with gate voltage is attributed to a decrease in Coulomb scattering with increasing carrier density. The Hall mobilities in this figure are plotted as a function of temperature in Figure 4 (a) of the main paper.

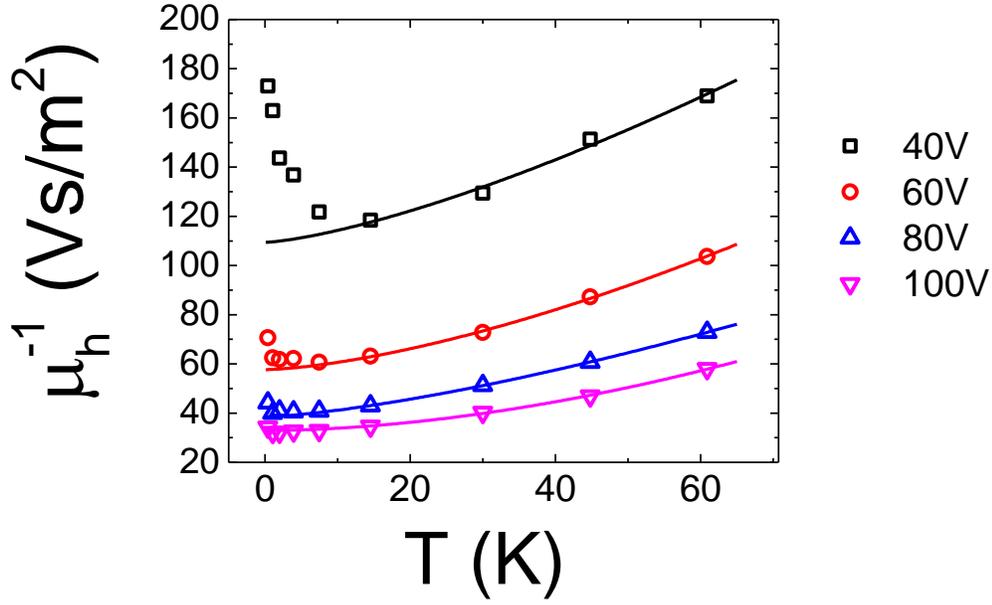

Figure S3: Reciprocal of the Hall mobility as a function of temperature for different back gate voltages, calculated from the mobilities plotted in Figure 4 (a) of the main paper. Symbols are the measured data while solid lines are fits to the data for T > 10 K. The data is fit using the following equation

$$\mu^{-1} = \mu_o^{-1} + AT^{\gamma} \tag{S1}$$

The characteristic exponent γ determined from the fitting above is plotted in Figure 4 (b) of the main paper.

**Hall Mobility in MoS$_2$, 110 K and Room Temperature**

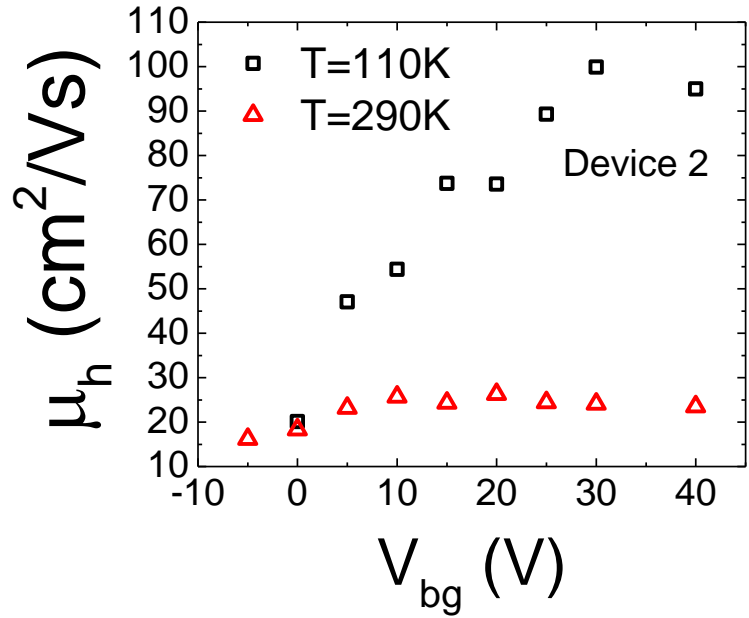

Figure S4: Hall mobility of multilayer MoS$_2$ as a function of back gate voltage. The device is different than that discussed in Figures S2 and S3. The increase in mobility with gate voltage at 110 K is attributed to a decrease in Coulomb scattering with increasing carrier density, just as in Figure S2. At room temperature, the mobility quickly saturates as a function of gate bias as shown in the figure. Hall mobility as a function of temperature for this device is plotted in Figure 5 (a) of the main paper.

**Thickness Characterization of MoS$_2$ devices**

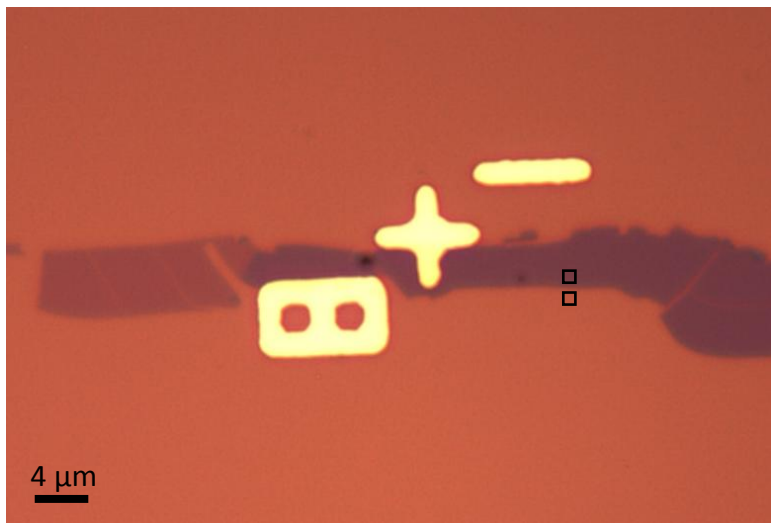

Figure S5: Optical image of the flake for Device 1. The relative optical contrast of the flake for the red channel is determined to be 0.36. This relative contrast was computed using the average red channel values within the boxes shown in the figure. The determined relative contrast indicates that the flake is four layers thick according to reference (S1). We note that that (S1) also used a 300 nm SiO$_2$ substrate, the same used in this work, so we can directly compare to their relative contrast values.

Reference

(S1)  Liu, Y.; Nan, H.; Wu, X.; Pan, W.; Wang, W.; Bai, J.; Zhao, W.; Sun, L.; Wang, X.; Ni, Z. Layer-by-Layer Thinning of MoS$_2$ by Plasma. *ACS Nano* **2013**, *7*, 4202–4209.

(a)

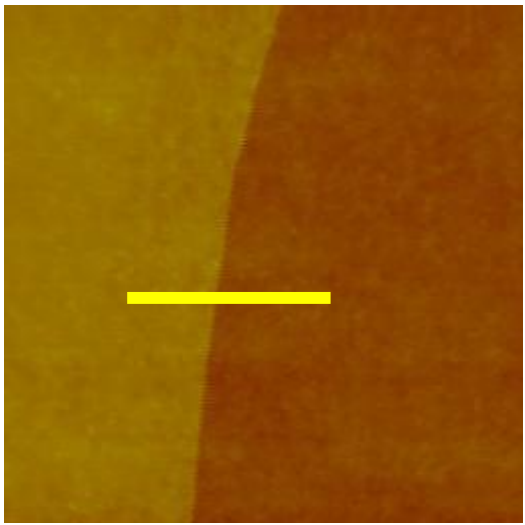 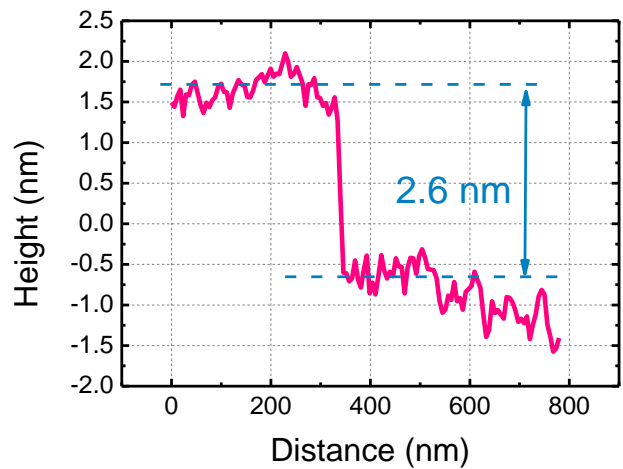

(b)

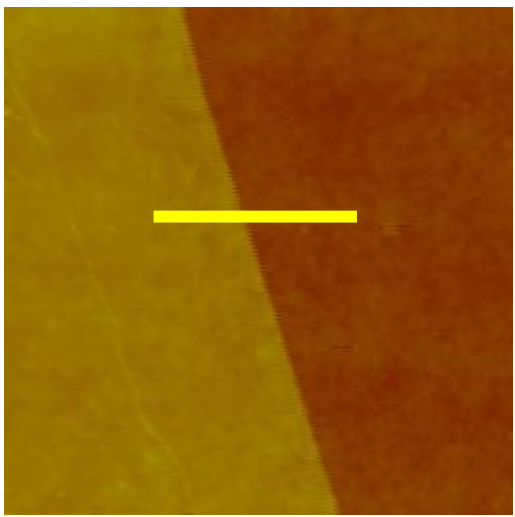 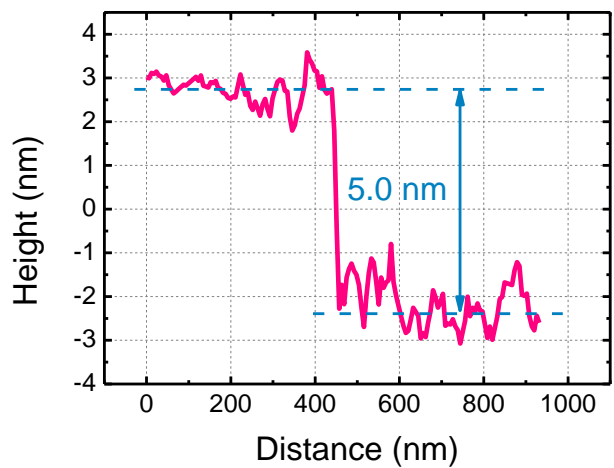

Figure S6: (a) AFM of the MoS$_2$ flake of the Device 1 which confirms the conclusion of Figure S5. (b) AFM of the similar MoS$_2$ flakes made for Device 2 and Device 3 with the thickness of 5-6nm.